\documentclass[apj]{emulateapj}
\begin{document}

\newcommand {\peryr}{$\rm{yr}^{-1}$}
\newcommand {\permyr}{$\rm{Myr}^{-1}$}
\newcommand{\nsns}{NS$-$NS}
\newcommand{\nswd}{NS$-$WD}
\newcommand{\wdns}{WD$-$NS}
\newcommand{\tlife}{$\tau_{\rm life}$}
\newcommand{\tmrg}{$\tau_{\rm mrg}$}
\newcommand{\td}{$\tau_{\rm d}$}
\newcommand{\tc}{$\tau_{\rm c}$}
\newcommand{\tsd}{$\tau_{\rm sd}$}
\newcommand{\rate}{$\cal R$}
\newcommand{\rtot}{${\cal R}_{\rm tot}$}
\newcommand{\rpeak}{${\cal R}_{\rm peak}$}
\newcommand{\rdet}{${\cal R}_{\rm det}$}
\newcommand{\ntot}{$N_{\rm tot}$}
\newcommand{\npsr}{$N_{\rm PSR}$}
\newcommand{\nobs}{$N_{\rm obs}$}
\newcommand{\nmean}{$<N_{\rm obs}>$}
\newcommand{\solarM}{M$_{\rm \odot}$}
\newcommand{\chmass}{$\cal M$}
\newcommand{\fmin}{$f_{\rm min}$}
\newcommand{\fmax}{$f_{\rm max}$}

\title{Constraining population synthesis models via 
  the binary neutron star population}
\author{R.\ O'Shaughnessy,  C.\ Kim, T.\ Fragos, V.\ Kalogera, \& K.\ Belczynski}

\begin{abstract}
The observed sample of double neutron-star (NS-NS) binaries presents a challenge to
population-synthesis models of compact object formation: the
input model parameters must be carefully chosen so the results match (i) the observed star formation rate and (ii) the formation rate of
NS-NS binaries, which can be estimated from the observed sample and
the selection effects related to the discoveries with radio-pulsar surveys.  In this paper, we
select from an extremely broad family of possible population synthesis
models those few (2\%) which are consistent with the rate implications of the observed sample of NS-NS binaries.
To further sharpen the constraints the observed NS-NS population
places upon our understanding of compact-object formation processes,
we separate the observed NS-NS population into two channels: (i)
merging NS-NS binaries, which will inspiral and merge
through the action of gravitational waves within 
 $10\,$~Gyr, and (ii) wide NS-NS binaries,
consisting of all the rest.   With the  subset of astrophysically
consistent models, we explore the implications for the rates at which double black hole (BH-BH), black hole-neutron star (BH-NS), and NS-NS
binaries merge through the emission of gravitational waves. 
\end{abstract}

 
\maketitle

\section{Introduction}
Interest in the formation channels and rates of double compact objects
(DCOs) has increased in recent years partly because, at the late stages of their inspiral through the emission of gravitational waves, they can be strong enough sources to be detected by the many presently-operating ground-based gravitational-wave detectors (i.e., LIGO, GEO, TAMA).
But with the notable
exception of NS-NS mergers -- see
Kim, Kalogera, and Lorimer (2003),
henceforth denoted KKL -- the merger
rates for DCOs with black holes  have not been 
constrained empirically.  
The only route to rate estimates for double black hole (BH-BH) and
black hole-neutron star (BH-NS) binaries  is through population
synthesis models. These involve a Monte Carlo exploration of the likely
life histories of binary stars, given 
statistics governing the initial conditions for binaries and 
a method for following the behavior of single and
binary stars 
\citep[see, e.g.,][]{StarTrack}.
Unfortunately, our understanding of the evolution of single and binary
stars is incomplete, and we parameterize that uncertainty with a great
many parameters ($\sim 30$), many of which can cause the predicted
DCO merger rates to vary by more than an order of
magnitude when varied independently through their plausible range.
To arrive at more definitive answers for DCO merger rates,
we must substantially reduce our uncertainty in the parameters that
enter into population synthesis calculations through comparison with observations.

Clues  to the physics 
underlying the formation of tight compact binaries
can be obtained through a study of each individual known DCO system.
Some authors have followed this path, for example 
examining the potential evolutionary and
kinematic histories of each individual binary to deduce the 
pulsar kicks needed to reproduce their evolutionary path
\citep[e.g.,][]{Bart1}.
However, the simplest and most direct way to constrain the parameters
of a given population synthesis code is to compare several of  its
many predictions against observations.
For example, the empirically estimated formation rates derived from the six known Galactic
NS-NS binaries --  half of which are tight enough to merge through the emission of
gravitational waves within $10$\,Gyr -- should  be reproduced by any
physically reasonable combination of model parameters for population synthesis.

In this paper, we describe the constraints the observed NS-NS
population places upon the most significant parameters that enter into
one population synthesis code, \emph{StarTrack} \citep{StarTrack, StarTrackUpdates}.
Furthermore, we use the set of models consistent with this
observational constraint to revise our population-synthesis-based
expectations for various DCO merger rates.

In Section \ref{sec:data} we describe the observational constraints from NS-NS: 
reviewing and extending the work of Kim, Kalogera, and Lorimer (2003)
we briefly summarize the observed sample of
NS-NS binaries, the surveys which detected them, and the implications
of the (known) survey selection effects for the expected
NS-NS formation rate.   In Section \ref{sec:ps} we
describe our population synthesis models and their predictions for
NS-NS  binary (and other compact binary) formation rates.  
Since a comprehensive population
synthesis survey of all possible models is not computationally
feasible, we describe an efficient approximate fitting technique
[used previously in 
O'Shaughnessy, Kalogera, and Belzcynski (2005); hereafter OKB] we
developed to accurately approximate the results of complete population synthesis
calculations.
Finally, in Section
\ref{sec:constraints} we select from our family of possible models
those predictions which are consistent with the observational
constraints.  We then employ that sample to generate refined
predictions for the expected BH-BH, BH-NS, and NS-NS merger rate
through the emission of gravitational waves.

We find the observed NS-NS population can provide a  tight
constraint (albeit a complicated one to interpret) on the many
parameters entering into population synthesis models.  With this
article serving as an outline of the general method, we propose to
impose in the future several additional constraints, including notably
the lack of any observed BH-NS systems, the empirical supernova rates
as well as formation rates of binary pulsars with white dwarf
companions.

\section{Empirical rate constraints from the NS-NS Galactic sample}\label{sec:data}

Seven NS-NS binaries  have
been discovered so far in the Galactic disk. Recently KKL developed a statistical method 
to calculate the probability distribution of rate estimates derived
using the observed sample and modeling of survey
selection effects. Four of the known systems will have merged within $10$\,Gyr (i.e., ``merging'' binaries: PSRs J0737-3039, B1913+16, B1534+12, and J1756-2251) and three are
wide with much longer merger times (PSRs J1811-1736, J1518+4904, and J1829+2456). PSR~J1756-2251 was discovered recently \citep{NS11} and has not been included in the current calculations. 
We do not expect, however, that this system will
significantly change our expectations of the merger rate. (see \citet{Kalogera2004} for details). This fourth system is
sufficiently similar to
PSR~B1913+16 and was discovered with pulsar 
acceleration searches, the selection effects of  which have already been accounted for (see Table~\ref{tab:NS} for the properties of the six systems used here).

In what follows we use observational constraints based on the rate
probability distribution derived by 
\citet{Kalogera2004,Kalogera2004b}  
for
merging binaries and the equivalent results for the three wide
binaries (presented for the first time in this article).  In
what follows we use the index $k=1\ldots 6$ to refer to the three
merging (1,2,3) and the three wide (4,5,6) NS-NS binaries.

\subsection{Merging NS-NS Binaries}

The discovery of NS-NS binaries with radio pulsar searches and our
understanding of the selection effects involved allows us to estimate the
total number of such systems in our Galaxy and their formation
rate. KKL developed  a statistical analysis designed to account
for the small number of known systems and associated
uncertainties. 
Specifically, they found that the posterior probability distribution
function ${\cal P}_k$ for NS-NS formation rates ${\cal R}_k$ for each
sub-population $k$ of pulsars similar to the $k$th known binary pulsar is given by 
 \begin{equation}
 \label{eq:coreDistribution}
  {\cal P}_k({\cal R}) = A_k^2 {\cal R} e^{-A_k {\cal R}}.
 \end{equation}
The parameter $A_k$ depends on some of the properties of the pulsars
in the observed  NS-NS  sample [see KKL Eq.~(17)]:
\begin{equation}
\label{eq:A}
A  = \tau_{life}/(f_b N_{PSR})
\end{equation}
where $f_b^{-1}$ is the fraction of
all solid angle the pulsar beam subtends; $\tau_{life}$ is the 
total binary pulsar lifetime
\begin{equation}
\label{eq:tau:merging}
\tau_{life} = \tau_{sd} + \tau_{mrg}  \qquad \text{(merging)}
\end{equation}
(where $\tau_{sd}$ is the pulsar spindown age [see Arzoumanian, Cordes, \& Wasserman 1999] and $\tau_{mrg}$ is the time
remaining until the pulsar merges through the emission of
gravitational waves [see Peters 1964 and Peters and Mathews 1963]);
and $N_{PSR}$ is the total estimated number of systems similar to each
of the observed one (i.e., $N_{PSR}^{-1}$ is effectively a volume-weighted probability that a
pulsar with the same orbit and an optimally oriented beam would be
seen with a conventional survey; this factor incorporates all our knowledge of pulsar survey selection
effects as well as the pulsar space and luminosity  distributions). 
Table~\ref{tab:NS} lists for each
merging NS-NS binary several intrinsic parameters (i.e., the best
known values for $f_b$; several lifetime-related parameters, such as
$\tau_{sd}$ and $\tau_{mgr}$) and two key quantities which depend on
our analysis of selection effects: $N_{PSR}$ and the deduced $A$
[i.e., via
Eq.~(\ref{eq:tau:merging})].  [Results are shown for our
preferred model for binary pulsar space and luminosity distribution; see model \#\,6 and details in KKL.]
 The total NS-NS posterior density of the
combined rate represented by the observed samples can be computed by a
straightforward convolution, 
\begin{eqnarray}
{\cal P}({\cal R}_{tot}) &=& \int d{\cal R}_1 d{\cal R}_3 d{\cal R}_2
\delta ({\cal R}_{tot} - {\cal R}_1 -{\cal R}_2 -{\cal R}_3) \nonumber
\\
 &\times&  {\cal P}_1({\cal R}_1) {\cal P}_2({\cal R}_2) {\cal P}_3({\cal R}_3)
\end{eqnarray}
described in detail in Section 5.2 of KKL
and presented in detail for the three-binary case in Eq.~(A8) of
\citet{Kim2004b}. 

KKL also demonstrated that the resulting rate distributions depend only weakly on the spatial distribution of NS-NS locations (see their
Figure 7 and the end of their Section 6).  Thus the NS-NS
rate distribution effectively depends on only one model
assumption,  the choice of the intrinsic radio pulsar luminosity
function -- which, in the KKL approach 
is given by [see KKL Eq.~(3), following Cordes and Chernoff (1997)],
\begin{equation}
\phi(L) dL = (p-1)(L/L_{\text{min}})^{-p} dL/L_{\text{min}}. 
\end{equation}
 Thus it is controlled by two parameters, the minimum allowed pulsar 
luminosity ($L_{min}$) and the power law $p>1$ governing their
relative luminosity probabilities.

KKL did not complete their calculation for a {\em comprehensive} posterior
probability distribution for the NS-NS rate estimates, however, because
up-to-date empirical probability constraints for $p$ and
$L_{min}$ are not available (cf., \cite{Proc}). Instead, they presented results for a few selected models,
emphasizing one model (model 6) whose properties
($L_{min}=0.3$mJy\,(kpc)$^2$ and $p=2$) are close to the median values they expect will be
found when all present observations are taken into account.
For this particular model, the empirical parameters $A_k$ which
describe the posterior densities are given in Table~\ref{tab:NS}.

\begin{deluxetable}{lrrrrrrrrrrrrc}[ht]
\tablecolumns{14}
\tablewidth{490pc}
\tablecaption{Observational properties of NS-NS binaries}
\tabletypesize{\footnotesize}
\tablehead{
\colhead{PSRs} &
\colhead{$P_{\rm s}^{a}$} &
\colhead{{\it \.{P}}$_{\rm s}^{b}$} &
\colhead{$P_{\rm b}^{c}$ }&
\colhead{M$_{\rm c}^{d}$} &
\colhead{e$^{e}$} &
\colhead{$\tau_{\rm c}^{f}$} &
\colhead{$\tau_{\rm sd}^{g}$} &
\colhead{$\tau_{\rm mrg}^{h}$} &
\colhead{$\tau_{\rm d}^{i}$} &
\colhead{N$_{\rm PSR}^{j}$} &
\colhead{$f_b^{k}$}& 
\colhead{$A^{l}$}& 
\colhead{Refs$^{m}$}
 \\ 
\colhead{} &  \colhead{(ms)} &  \colhead{($10^{-18}$s~s$^{-1}$)} &  \colhead{(hr)} & \colhead{(M$_{\rm \odot}$)} &
\colhead{} & \colhead{(Gyr)} & \colhead{(Gyr)} &  \colhead{(Gyr)} & \colhead{(Gyr)} &  \colhead{} & \colhead{} &\colhead{(Myr)} & \colhead{}
}
\startdata
(1) merging NS-NS & &&&&&&&&&&&&\\
\hline
B1913+16 & 59.03 & 8.63 & 7.752 & 1.39 & 0.617 & 0.11 & 0.065 & 0.3 & 4.34 & 617 & 5.72 & 0.103  & 6,7\\
B1534+12 & 37.90  & 2.43 & 10.098  & 1.35 & 0.274 & 0.25 & 0.19 & 2.7 & 9.55 & 443 & 6.45 & 1.014 & 8,9\\
J0737-3039 & 22.70  & 1.74 & 2.454  & 1.25 & 0.088 & 0.16 & 0.10 & 0.085 & 13.5 & 1621 & 6.085 & 0.018 & 10\\
\hline
(2) wide NS-NS& &&&&&&&&&&&&\\
\hline
J1811-1736 
& 104.182 & 0.916     & 450.7  & 1.66 & 0.828 & 1.8 & 1.8 & n/a & 7.8 & 606 & 6 & 2.64 & 13 \\
J1518+4904        & 40.935  & 0.02        & 207.216& 1.35 & 0.25  &
32.4 & 32.3 & n/a & 54.2 & 282 & 6 &32.9 & 14,15 \\
J1829+2456        & 41.0098 & $\sim$ 0.05 & 28.0   &
1.15 & 0.139 & 13.0 & 12.9 & n/a & 43.7 & 272 & 6 & 37.9 & 16 \\
\enddata
\label{tab:NS}
\tablenotetext{a}{Spin period.}
\tablenotetext{b}{Spin-down rate.}
\tablenotetext{c}{Orbital period.}
\tablenotetext{d}{Estimated mass of a companion, where M$_{\rm NS}$ is assumed to be 1.35 M$_{\rm \sun}$. PSR J1141$-$6545 has M$_{\rm NS}$=1.30 M$_{\rm \sun}$. M$_{\rm NS}$=1.44 (B1913+16) and 1.33 M$_{\rm \odot}$ (B1534+12).}
\tablenotetext{e}{Eccentricity.}
\tablenotetext{f}{Characteristic age of a pulsar.}
\tablenotetext{g}{Spin-down age of a pulsar. We calculate $\tau_{\rm sd}$ only for those recycled pulsar.}
\tablenotetext{h}{Merging time of a binary system due to the emission of gravitational waves.}
\tablenotetext{i}{Death time of a pulsar.}
\tablenotetext{j}{Most probable value of the total number of pulsars in a model galaxy estimated for the reference model (model 6 in KKL).}
\tablenotetext{k}{Beaming factor for pulsar}
\tablenotetext{l}{Parameter in rate equation [see Eq.~(\ref{eq:A})].}
\tablenotetext{m}{References: 
(1) Lundgren, Zepka, \& Cordes (1995);
(2) Edwards, \& Bailes (2001) ; 
(3) Kaspi et al.\ (2000) ;
(4) Bailes et al. (2003) ;
(5) van Kerkwijk, \& Kulkarni (1999); 
(6) Hulse \& Talor (1975); 
(7) Wex, Kalogera, \& Kramer (2001); 
(8) Wolszczan (1991); 
(9) Stairs et al.\ (2002)
(10) Burgay et al.\ (2003)
(11) Faulkner et al. (2004)
(12) Lyne et al. (2000)
(13) Corongiu et al. (2004)
(14) Nice, Sayer, \& Taylor (1996)
(15) Hobbs et al. (2004)
(16) Champion et al. (2004)
}
\end{deluxetable}

\subsection{Wide NS-NS Binaries}
The same general technique outlined above can be applied to the
formation rate of wide NS-NS binaries: the same form of distribution function 
${\cal P}({\cal R})$ [Eq.~\ref{eq:coreDistribution}] applies and it depends on the same parameter $A$ [Eq.~\ref{eq:A}].
The main change is the relevant lifetime.  Since these binaries do not
merge, their detectable lifetime is now the sum of the time remaining before the pulsar spins down ($\tau_{sd}$, described earlier) and the length of time the pulsar will remain visible (the ``death time'' of a
pulsar; see Chen \& Ruderman 1993).  However, since $\tau_{sd}$ estimates are somewhat uncertain, we require that they do not exceed the current age of the Galactic disk ($10$\,Gyr). To summarize, then, the only change from the previoius
approach is to replace the previous expression for the lifetime,
Eq.~(\ref{eq:tau:merging}), with
\begin{equation}
\label{eq:tau:wide}
\tau_{life}  =  \min{}(\tau_{sd}, 10 Gyr) +  \tau_d
\text{(wide)}  \; .
\end{equation}
Table~\ref{tab:NS} lists pulsar parameters and deduced quantities for
the three wide NS-NS binaries used in this study.
Current pulsar observations do not provide us with any estimates of
the beaming fractions relevant to the pulsars in these wide
systems. Guided by the beaming fraction distribution for  merging pulsars,
we adopt a value of 6 for the
beaming factor for all the wide NS-NS pulsars. 
For simplicity, we present the results for only the
preferred luminosity model (i.e., for the specific choice for $p$ and $L_{min}$ mentioned above). 
These distributions again follow Eq. (\ref{eq:coreDistribution}),
with parameters $A_k$ given by Table~\ref{tab:NS}, where $A_k$ is determined for each pulsar class $k$ 
from physical parameters presented in the table.  

\begin{figure}[h]
\includegraphics{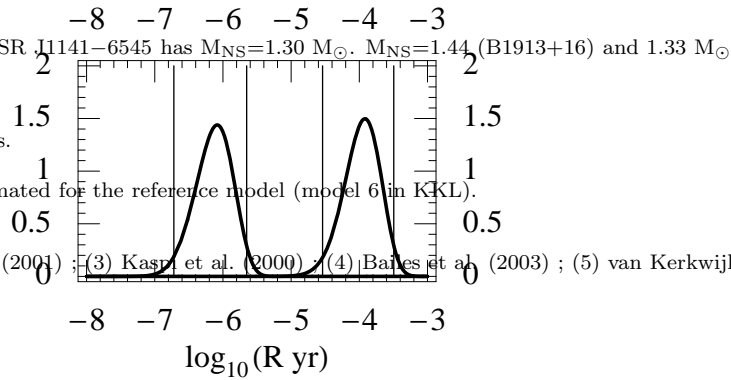}
\caption{\label{fig:ObservedNSRates}A plot of empirically-deduced
  probability distributions for merging (right) and wide
  (left) NS-NS binaries; see Sec.~\ref{sec:data}.  The solid vertical
  lines are at (i) $\log_{10}{\cal R}=-4.5388,-3.49477$, the 95\%
  confidence interval for the
  merging NS-NS merger rate;  and at (ii)  $\log_{10}{\cal
  R}=-6.7992,-5.7313$,
  the 95\% confidence interval for the wide NS-NS formation rate.
}
\end{figure}

For each class separately (merging and wide binaries) we
use  Eq.~(A8) of \citet{Kim2004b} to generate 
a composite probability distributions for the formation
rate of binaries in that class: ${\cal P}_m$ (merging) and ${\cal
  P}_w$ (wide).    
Thus we arrive at the two estimates shown in Fig.~\ref{fig:ObservedNSRates}
for the empirical probability distribution 
\[
p(\log {\cal R}) = {\cal P}({\cal R}) {\cal R} \ln 10
\] 
for the formation rates ${\cal R}_m$ and ${\cal R}_w$ of these two
classes of binary.  
>From these distributions we derive a 95\% confidence intervals for
each formation rate; for example, the upper and lower rate limits
${\cal R}_{w,\pm}$ satisfy
\begin{equation}
\int_0^{{\cal R}_{w,-}} d {\cal R} {\cal P}_w({\cal R}) =
\int_{{\cal R}_{w,+}}^\infty d {\cal R} {\cal P}_w({\cal R}) = 0.025
\; .
\end{equation}

\vspace*{5.cm} 

\section{Estimates for merger   rates}
\label{sec:ps}

\subsection{Population Synthesis Estimates}
We estimate formation and merger rates for several classes of double
compact objects using the 
\emph{StarTrack} code first developed by Belczynski, Kalogera, and Bulik
(2002) 
[hereafter BKB] and recently significantly updated and tested as described in detail in
Belczynski et al.\ 2005.  In this code, seven parameters strongly
influence compact object merger rates: the supernova kick distribution (3 parameters), the massive stellar wind strength (1), the common-envelope energy transfer efficiency (1), the fraction of mass accreted by the accretor in phases of non-conservative mass transfer (1), and the binary mass ratio distribution described by a negative power-law index (1).    To allow for an
extremely broad range of possible models, we used the specific
parameter ranges quoted in Section 2 of \citet{PSutility}.  

We randomly choose model parameters in this space
and evaluate their implications, by progressively  examining the
evolution of binary after binary.   We then extract from our
simulations predictions for several DCO formation rates (BH-BH, BH-NS,
and NS-NS) by scaling up the ratio of DCO formation events we obtain
in each simulation ($n$) to the total number of binaries studied in
the simulation ($N$) by a factor proportional to the expected ratio
between $N$ and the number of stars formed in the Milky Way.  We set this
scaling factor by assuming 
a constant star-formation rate of $\dot{M}\approx 3.5 M_\odot
\text{yr}^{-1}$, as described in the Appendix of OKB. 

\emph{Extracting predictions for the ``visible'' NS-NS formation rates}:
To compare the predictions of population synthesis calculations
against the empirical rate constraints derived for the pulsar samples,
we must determine the formation rates of NS-NS binaries that could be
``visible'' as pulsars. Since we do not follow the detailed pulsar
evolution with {\em StarTrack} (due to major uncertainties related to
pulsar magnetic field evolution), we choose a minimal criterion for
identifying NS-NS binaries that possibly contain a recycled pulsar: if
the first NS in the binary has experienced {\em any} accretion episode
(through either Roche-lobe overflow and disk accretion or a
common-envelope phase), then the binary is identified as a potential
binary recycled pulsar and is included in the calculation of the NS-NS
``visible''  pulsar formation rate.

\emph{Practical complications in merger rate calculations}:
We would have \emph{preferred} to proceed as in OKB and perform,
for each separate DCO type (e.g., BH-BH binaries), a sequence of
Monte Carlo computations tailored to determine this type's merger rate to some
fixed accuracy (say, 30\%) as a function of all population synthesis parameters.
Instead, owing to computational limitations, we had to extract
multiple types of information from each population synthesis run;
Appendix~\ref{ap:runs} describes in greater detail the collection of
population synthesis runs we performed and the manner in which these
runs were used to estimate various DCO formation rates.


\subsection{Mapping population synthesis rates versus parameters}
In order to constrain population synthesis parameters based on rate
measurements, we must be able to \emph{invert} the relation between
rate and model parameters to find all possible models consistent with
a given rate.   In other words, we \emph{must} fit the rates over all
seven parameters.

\begin{figure}
\includegraphics{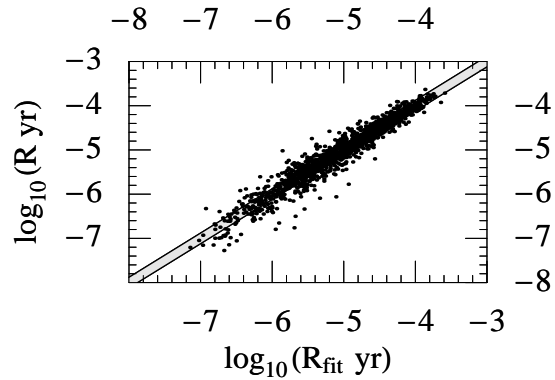}
\caption{\label{fig:rateSkew} 
This plot of the $\log_{10}$ of the Galactic rate versus our fit to the rate, shown for BH-BH, BH-NS, and NS-NS sample points (all superimposed). The shaded region is offset by a factor $1\pm 1/\sqrt{10}$.  This region estimates the error expected due to random fluctuations in the number of binary merger events seen in a given sample.   (See the appendix for a discussion of the number of sample points actually present in various runs.)
}
\end{figure}

\begin{figure}
\includegraphics{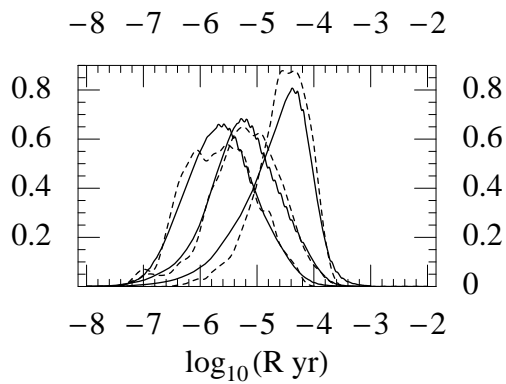}
\caption{\label{fig:histogram}
A plot of the a priori probability distribution for the 
BH-BH (left), BH-NS (center), and NS-NS (right) merger rates, versus the
$\log_{10}$ of the rate.  These distributions were generated from the
population synthesis code (dashed line) and fits (solid lines) assuming
all parameters in the population synthesis code were chosen at random
in the allowed region.
}
\end{figure}

OKB first demonstrated that, even using sparse data in a
high-dimensional space of population synthesis parameters, an
effective fit could be found for  formation rates of DCOs (see OKB
Fig. 2 and their Section 4).
We constructed separate polynomial  least-squares fits to each
of the five rate functions we need (i.e., for BH-BH,   BH-NS, visible
merging NS-NS, and visible wide NS-NS binaries we performed a cubic
least-squares fit; and for the overall NS-NS merger rate -- including
all merging NS-NS binaries, whether we  expect them to be
electromagnetically visible or not --  we used a quartic least squares 
fit).  Figure~\ref{fig:rateSkew}
demonstrates that the fit is good: the errors are on the limiting
scale we would 
expect, given the uncertainties in the input [i.e., the standard
deviation of the logarithmic rate errors,
$\log_{10} {\cal R}_{fit}/{\cal R}_{true}$ are $0.167$ (BH-BH),
$0.22$ (BH-NS), and
$0.086$ (NS-NS), are  comparable with the minimum possible uncertainty
we would expect given a perfect fit, $\log_{10}(1+1/\sqrt{10})\approx
0.119$; see Appendix \ref{ap:runs} for a detailed discussion of the
minimal uncertainties expected for each rate].   The fits are
sufficiently good that for our purposes we can  
replace population synthesis calculations with evaluations of our fits.
In particular and by way of example, in Figure \ref{fig:histogram} we
generate a histogram for various DCO formation rates predicted by
population synthesis, using 
(i) the actual outputs deduced from the population synthesis code, sampled at
random Monte Carlo points (dashed line), and (ii) the outputs obtained from fits to
the  dataset from (i), sampled at a much larger number of data points
(to insure smoothness; solid line).  The two methods produce strikingly similar
histograms, demonstrating that the fit will be adequate for our purposes.

\section{Constraints from NS-NS observations}
\label{sec:constraints}
In Sec.~\ref{sec:data} we constructed two straightforward empirical
constraints (i.e., confidence intervals for the formation rate of
``visible'' merging and wide NS-NS binaries)  we could place
on the output of population synthesis calculations.    In this
section, we apply these two constraints, individually and
together, and determine their effect on rate predictions of
population synthesis calculations.

\begin{figure}
\includegraphics{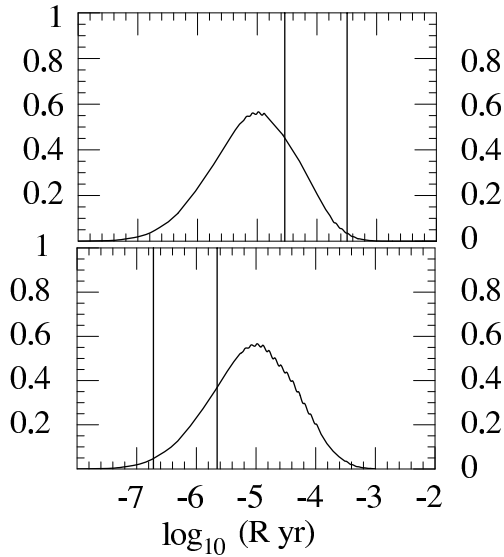}
\caption{\label{fig:theConstraints}
A plot of the a priori probability distributions for the visible merging (top)
and visible wide (bottom) NS-NS formation rates produced from population
synthesis.  The dashed curves  denote the direct result from population
synthesis; the solid curves denotes the result deduced from artificial
data generated from a multidimensional fit to the visible wide and
merging NS-NS rate data.  The vertical lines are the respective 95\%
CI bounds presented in Fig.~\ref{fig:ObservedNSRates}. 
}
\end{figure}

\emph{Bounding merging NS-NS rate}: Figure \ref{fig:theConstraints}
superimposes the observational bounds taken from the observed merging
NS-NS distribution
(i.e., the 95\% confidence interval; see Fig.~\ref{fig:ObservedNSRates}) on top of the distribution of
``visible'' NS-NS merger rates obtained from unconstrained population
synthesis.   
We limit attention only to
those
population synthesis models consistent with our constraint;
specifically, we randomly 
choose population synthesis models, evaluate the ``visible'' merging
NS-NS rate using our fit, and retain the model only if the rate lies
within these two bounds.
We find we reject 72\% of models we initially considered plausible.
For each of the small residual of consistent models, we can evaluate
the BH-BH, BH-NS, and NS-NS merger rates (again using our fits).  
We find the merger rates increase slightly on average: the mean merger
rate increases 
by a factor $\times 1.2$ for BH-BH, $\times 2$ for BH-NS, and $\times 2.2$ for NS-NS. 

\emph{Bounding wide NS-NS rate}: Figure \ref{fig:theConstraints}
also shows the 95\% confidence interval for the \emph{wide} 
visible NS-NS formation rate (Fig.~\ref{fig:ObservedNSRates}) on top of our a priori  population synthesis  distribution for  the wide visible NS-NS formation rate. We find that with this constraint we have to exclude 80\% of a priori plausible models. Using only models which satisfy this second constraint, we find DCO merger rates have dropped relative to our a priori predictions: the average BH-BH, BH-NS, and NS-NS merger rates   are reduced by a factor $0.65$, $0.26$, and $0.22$, respectively.

\emph{Both constraints simultaneously}: Very few population synthesis
models (less than $2\%$) satisfy both constraints simultaneously.  
Since the set of consistent models is much smaller than initially
permitted a priori, we have less uncertainty in our predictions:
the standard deviation in the log of the merger rates has changed from our initial a priori uncertainty of
0.60, 0.55, and 0.63
(i.e.,
plus or minus a factor of 
4, 3.5, and 4.3) 
for BH-BH, NS-NS, and BH-NS mergers, respectively, to 
0.61, 0.44, and 0.48 
(i.e., plus or minus a factor of 
4, 2.8, and 3).  
Further, since the wide constraint proves slightly more restrictive,
the  mean merger rates have dropped slightly from our prior expectations:
the average predicted merger rates are
1.1/Myr [BH-BH] (down by a factor $0.43$), 
1.4/Myr [BH-NS] (down by a factor $0.24$), and
6.7/Myr [NS-NS] (down by a factor $0.25$).

\begin{figure}
\includegraphics{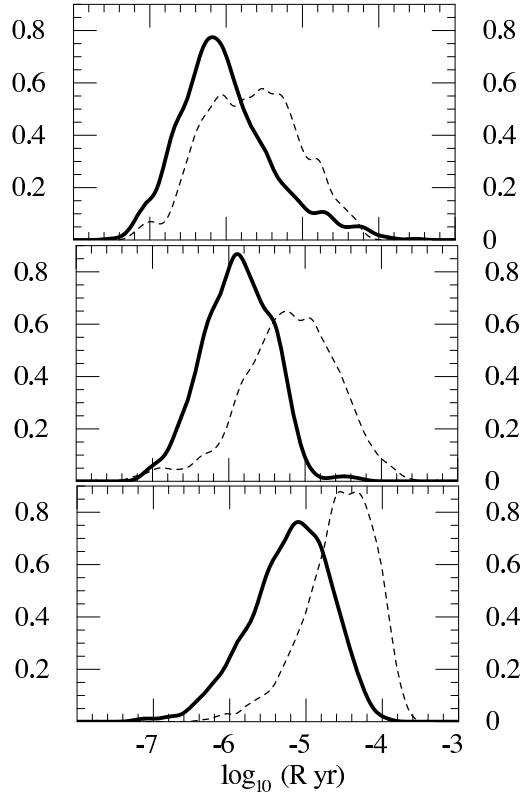}
\caption{\label{fig:AllConstraints}
A plot of the a priori probability distribution for the BH-BH
(top), BH-NS (center), and NS-NS (bottom) merger rates per
milky way equivalent galaxy.  As in Fig.~\ref{fig:histogram}, the dashed curves show the results 
obtained from our
population synthesis calculations (i.e., our raw code results, smoothed);
the thick solid curves show the results after we impose both our
observational constraints (i.e., consistency with the observed number of
visible wide and visible merging NS-NS binaries). 
}
\end{figure}

\subsection{Advanced LIGO detection rates}
While we have presented the number of mergers occurring \emph{per
Milky Way equivalent galaxy}, advanced LIGO's inspiral detection range depends on the masses of the component objects.  Specifically,
{\em one 4\,km advanced LIGO detector} (see Harry 2005) is expected to detect a binary with chirp mass 
${\cal M}_c = (m_1 m_2)^{3/5}/(m_1+m_2)^{1/5}$ (at signal-to-noise ratio 8) out to a
distance
\begin{equation}
d =  191 \, {\rm Mpc} \; \left({\cal M}_c/ 1.2 M_\odot \right)^{5/6}.
\end{equation}
Thus, if the real merger rate and chirp mass distribution 
for type $\alpha$ are ${\cal R}_\alpha$ and $p_\alpha({\cal M}_c)$, 
respectively, then LIGO will on average detect $\alpha$ merger events
at a rate
\begin{equation}
\label{eq:LIGORate}
{\cal R}_{\alpha,\rm LIGO} = 0.042 \; {\cal R}_\alpha \left< ({\cal M}_c/M_\odot)^{15/6} \right>_\alpha
\end{equation}
where 
$\left< ({\cal M}_c)^{15/6} \right>_\alpha = \int d{\cal M}_c p_\alpha({\cal M}_c) {\cal M}_c^{15/6}
$, and
where for simplicity we assume a uniform distribution of Milky Way
equivalent galaxies with density $0.01$/(Mpc)$^3$
(see \citet{Nutzman} for a discussion of short-scale corrections to this
distribution in the case of short-range interferometers, like initial LIGO).

\begin{figure}
\includegraphics{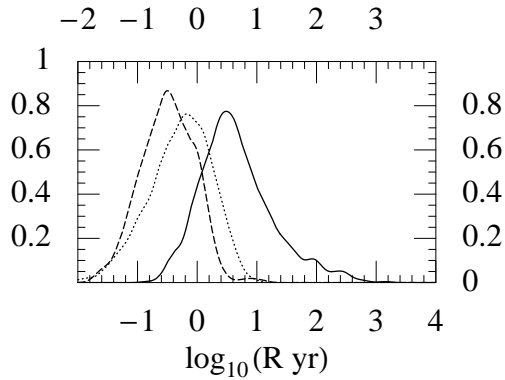}
\caption{\label{fig:LIGORates}
This plot shows our expectations for LIGO's  detection rates for merging
BH-BH (solid line), BH-NS (dashed line), and NS-NS (dotted line) binaries.
This plot was obtained directly from Fig.~\ref{fig:AllConstraints} using
Eq.~(\ref{eq:LIGORate}).
}
\end{figure}

Since  Fig.~\ref{fig:AllConstraints} was produced using  merger rate
fits,  
we do not have the chirp mass information needed to translate
that figure into a corresponding distribution for the LIGO detection rate.
In what follows we adopt mean chirp masses as derived from all our models in the archives. We find: 
 \begin{eqnarray} 
 & & \left< ({\cal M}_c)^{15/6} \right>_{\rm BH-BH} = 111 M_\odot^{15/6}
  \\
 & & \left< ({\cal M}_c)^{15/6} \right>_{\rm BH-NS} =  5.8 M_\odot^{15/6}
 \\
 & & \left< ({\cal M}_c)^{15/6} \right>_{\rm NS-NS} =  2 M_\odot^{15/6}.
 \end{eqnarray} 
 These are to be compared to $224 M_\odot^{15/6}$ for two $10$\,M$_{\odot}$ BH, to $15.5 M_\odot^{15/6}$ for a $10$\,M$_{\odot}$ BH and a $1.4$\,M$_{\odot}$ NS, and to $1.2 M_\odot$ for two $1.4$\,M$_{\odot}$ NS.
Figure~\ref{fig:LIGORates} presents our preliminary estimates for the
advanced LIGO detection rate distribution.  Note this figure 
provides the same
information as Figure~\ref{fig:AllConstraints}, except that the merger
rates for each species have been rescaled according to
Eq.~(\ref{eq:LIGORate}).

\section{Summary and conclusions}
In the context of matching theory with observations of the binary NS
population, we have described  how to constrain the predictions for DCO merger rates population synthesis
codes such as \emph{StarTrack} by using two specific observational
constraints.  We find that to be consistent with the rate statistics of the
observed NS-NS population (at the 95\% confidence interval), we must exclude at least 98\% of
all the models we think a priori likely. We do not focus on explicitly describing the seven-dimensional region of {\em StarTrack}  model
parameters consistent with our constraints, both because (i) we lack a
compact way to describe a seven-dimensional region, and (ii) our
region has meaning specifically for the StarTrack code.  Other codes
have different parameterizations of the same physical phenomena,
leading to potentially quite different 
representations of the same constraint region.
However, in Appendix~\ref{ap:describeregion} we give some information about the mean constraints on population synthesis parameters but also the strong variance around these mean values. 
As described in Section \ref{sec:constraints} and
particularly via Fig.~\ref{fig:AllConstraints}, to extract a
physically meaningful statement about the  
effect of imposing constraints (as opposed to a 
describing information about parameters of one particular code), 
we have described how these
constraints have improved our understanding of three DCO merger rates
[BH-BH, BH-NS, and NS-NS]. 
We find that, using these two initial constraints, (i) the most probable
merger rates (i.e., at peak probability density) are systematically
lower than we would expect a priori, at least by a factor 2; and (ii) 
we reduce the uncertainty in the BH-NS and NS-NS merger rates by
moderate factors (i.e., the standard deviation of $\log{\cal R}$ drops
by 0.03 and 0.16, respectively).

This paper only outlines the beginning of a large program we have undertaken to better constrain our understanding of the evolution of
single and binary stars and the associated predictions for
gravitational-wave sources.  We intend to add a few additional
empirical constraints of DCOs (e.g., WD-NS binaries) and the lack of
observations of certain binary compact objects (notably, BH-NS
binaries). Apart from rate constraints, the observed properties of
DCOs (mass ratios, orbital separations and eccentricities) could also
be used as constraints. Further 
constraints (such as observations of pulsar kicks, which constrain the
supernova kick distribution) can be added by other means, as prior
distributions on the space of model parameters.   We fully expect to
have much stronger constraints on our understanding of population
synthesis in the near future.

Stronger constraints, however, will require a considerably more
systematic approach than the straightforward presentation we have used
here.  As the expected uncertainties decrease, greater
care must be taken to include \emph{every} uncertainty, no matter how
minor, many of which for clarity we have neglected.  For example, in
future calculations we expect to include uncertainties in  $L_{min}$
and $p$, our fit 
and rate estimates, and even the star formation rate
we use to convert simulation results into rate estimates.
Additionally, we will  self-consistently choose the constraint confidence
intervals in order to construct a meaningful posterior confidence
interval on each merger rate.

\emph{Note on BH-pulsar rates}: Recently, \citet{Pfahl} have estimated
 that BH-millisecond pulsar systems should be exceedingly rare: they
 find an
 upper bound of $10^{-7}$/year on the formation rate of binary systems
containing a BH and a recycled pulsar.  In our own computations, we
have \emph{never} seen such a system form, even though we have seen
 $\approx 6\times 10^4$ merging NS-NS binaries form.  This result
suggests the branching ratio for BH-PSR:NS-NS formation is $\ll
 10^{-4}$.  If we use a conservative value for the merging NS-NS
 formation rate, $10^{-4}/{\rm yr}/{\rm galaxy}$, then we expect the
 formation rate of BH-PSR binaries to be significantly less than
 $10^{-8}$/yr/galaxy, entirely consistent with  their constraint. 

\acknowledgements

We thank David Champion for providing spin-down data for PSR~J1829+2456.  This work is partially supported by NSF Gravitational
Physics grants PHYS-0121416 and PHYS-0353111, a David and Lucile Packard Foundation Fellowship in Science and Engineering, and a Cottrell Scholar Award from the Research Corporation to VK.

\appendix

\section{A.~Calculating DCO event rates with population synthesis}
\label{ap:runs}
This paper relies upon formation rates extracted from a large sequence of
archived population synthesis calculations.   This appendix describes
how these archives were generated and used to produce formation
rates.  It also explains the expected uncertainties in each merger
rate estimate.

\subsection{How rates were estimated}

\emph{Principles of archive generation:}
For a given combination of population synthesis parameters, we
generate a large collection $N$ of binaries.  To
add a binary to the archive, we first generate progenitor binary
parameters ($m_1,m_2,a,e$) [i.e., the two progenitor masses
($m_{1,2}$) and the initial semimajor axis ($a$) and eccentricity
($e$)] which are (i) drawn from the
distribution functions presented in OKB and which (ii) satisfy any
conditions  we impose to reject binaries irrelevant to the study at
hand -- e.g., $m_{1,2}>4$, or more elaborate conditions described in
OKB.  (The latter conditions offer a significant speed improvement,
but rely upon the experience gained in previous runs to insure that
the conditions imposed do not reject physically relevant systems.)
Given satisfactory initial conditions, binaries are assigned a
randomly-chosen formation time, then evolved from whenever they form
until the present day.
Binaries are
successively added to the archive until some termination condition is reached
-- typically, that the number $n$ of a given class of binaries, such as
merging NS-NS binaries, has crossed a threshold (i.e., $n=10$).

\emph{Archive classes:}
Our population synthesis runs are 
summarized in Table~\ref{tab:runs}.  
Each run of the population synthesis code falls into a certain class,
depending on what choices were made for (i) the target systems on
which the termination threshold was set (i.e., stop when we get $n$
merging BH-BH binaries; see column 2 of Table~\ref{tab:runs}), (ii) the specific threshold $n$ chosen (i.e.,
which insures that the formation rate of the target system type is
determined to an accuracy roughly $\sim 1/\sqrt{n}$; see column 3 of
Table~\ref{tab:runs}), and (iii) the 
combination of conditions applied to filter progenitor binary systems
(see the last column of Table~\ref{tab:runs}).  
In Table~\ref{tab:runs}, the filters B, and
S correspond to using the partitions presented in listed in Table~2 of
OKB for BH-BH binaries (B) and NS-NS binaries (S); the 'W' filter uses
only the \emph{first} NS-NS partition listed in Table 2 of OKB, which
filters out WD progenitors.  Note the first column of Table~\ref{tab:runs} merely provides
a label for the archive class.
%


\emph{Applying archives}:
>From the ratio of the number of binaries of a given type seen to the
number of binaries in a run, modulo a normalization factor presented
in OKB, we can calculate the formation rates for any
binary type of interest.  However, to avoid extreme biases associated
with a poor choice of filter or stopping condition, we use only
certain archives to estimate merger rates, as described in
Table~\ref{tab:rates}.  [In this table, all rates are total merger
rates, 
with the exception of the last two rows, which correspond to the
visible merging (v) and visible wide (vw) NS-NS binaries.]

\begin{deluxetable}{llrrr}
\tablecolumns{5}
\tablecaption{Classes of runs}
\tablehead{\colhead{Type} & \colhead{Target} & \colhead{Number of
runs} & \colhead{n} & \colhead{Filters}}
\startdata%
a    & NS-NS & 488 & 10 & (none)\\
a'   & NS-NS & 137 & 100& S\\
a''  & NS-NS & 408 & 300& W\\
b    & BH-BH & 306 & 10 & B\\
b'   & BH-BH & 285 & 10 & W\\
c    & BH-NS & 357 & 10 & W%
\enddata
\label{tab:runs}
\end{deluxetable}

\begin{deluxetable}{l|cccccc}
\tablecolumns{7}
\tablecaption{Classes used for specific rates}
\tablehead{ 
  Type        & a & a' & a'' & b & b' & c}
\startdata
  BH-BH       &   &    &     & x &    &   \\
  BH-NS       &   &    & x   &   & x  & x \\
  NS-NS       &   & x  & x   &   & x  & x \\
  NS-NS(v)    & x & x  & x   &   & x  & x \\
  NS-NS(vw)   &   & x  & x   &   & x  & x
\enddata
\label{tab:rates}
\end{deluxetable}

\subsection{Understanding errors in rate estimates}
Tables \ref{tab:runs} and \ref{tab:rates} provide the information
needed to understand  errors in our formation rate
estimates.    

\emph{Example}: The BH-BH formation rate estimate is produced from
a single archive ($b$).  Archive $b$ is a collection of runs which stop when
10 merging BH-BH binaries are found; while the filters in this archive
can prevent the formation of binaries involving NS, they do not
significantly limit BH-BH binary formation.  Therefore, archive $b$ is
ideally suited to estimate the BH-BH merger rate to an accuracy of
order $1/\sqrt{10}\approx 30\%$.

\emph{Example}: The NS-NS formation rate estimate is produced from an
amalgam of archives ($a'$, $a''$, $b'$, and $c$).  None of these
archives applies filters which prevent NS-NS formation, though one
($a'$) applies filters which prevent the formation of nearly anything
else.  However, these archives do involve different termination
criteria: the first two terminate when a large number of NS-NS
binaries have formed, whereas the last two terminate when only 10
BH-BH or BH-NS archives have formed.  If only the first two were used,
we could guarantee the NS-NS merger rate to be known to within
$10\%$ accuracy.  However, to augment our statistics, we additionally included
binaries from $b'$ and $c$; while these archives \emph{should usually} have many
NS-NS binaries (cf. Fig.~\ref{fig:histogram}), we cannot guarantee any
minimum number a priori.    To simplify error estimates, we selected
only those elements of $b'$ and $c$ with more than $30$ merging NS-NS
binaries.  Thus, we expect our NS-NS rate to be known to within
$20\%$ accuracy.

\emph{Example}: The estimate for the visible wide NS-NS formation rate is the least accurate and most challenging calculation we performed.  Since visible wide NS-NS systems were rare and since we did not (unlike the BH-BH case) have a simulation dedicated to discovering them, we had to scavenge through all the archives which could have produced them
(i.e., not $b$) in sufficient numbers (i.e., not $a$) to permit a
moderately accurate rate estimate.    In practice, we selected those
runs which formed more than $8$ wide visible NS-NS binaries, which
should in principle give us an accuracy of order (few)$\times 35\%$.
[In practice, we found a roughly 75\% accuracy, comparable to the 65\%
accuracy of our next-most-accurate estimate (the BH-NS rate).]

\section{B.~Sample fits to merger rates}
\label{ap:samplefits}
This paper and in OKB rely upon fits to 7-dimensional functions
obtained from the \emph{StarTrack} population synthesis code.   In
this section, we provide an {\em example} of an explicit formula for a \emph{quadratic-order polynomial fit} to the BH-BH merger rate.  
We express this fit in terms of the following dimensionless
parameters $x_k\in[0,1]$:
$x_1= r/3$ where $r\in[0,3]$ is a negative power-law index describing the mass ratio
distribution  assumed for binaries; $x_2= w$ characterizes the strength of stellar winds; the kick velocity distribution consists of two Maxwellian distributions with 1-D dispersions of $\sigma_{1}$ and $\sigma_{2}$, which are varied within [0,200\,km\,s$^{-1}$] and (200, 1000\,km\,s$^{-1}$], respectively, using two parameters: 
$x_3= \sigma_1/(200\, \text{km/s})$ and 
$x_4= -1/4 +\sigma_2/800\, \text{km/s})$; a third parameter $x_5= s$ is used as the relative weight between low  and high kick magnitudes; 
 $x_6=\alpha \lambda$ is the effective common-envelope
efficiency; and $x_{7}=f_a$  is  the fraction of mass accreted by the accretor in phases of non-conservative mass transfer. 
[These parameters are discussed more thoroughly in OKB and in the
original \emph{StarTrack} paper \citet{StarTrack}.]  In terms of these
parameters, we find the following quadratic fit to the BH-BH merger rate: 
\begin{eqnarray}
\log_{10}\left[ {\cal R}_{\rm BH-BH} {\rm yr} \right] &=& 
-5.84517 + 1.30448\,{x_1} - 0.406066\,{{x_1}}^2 
\nonumber\\&&
 - 0.310686\,{x_2} - 0.407175\,{x_1}\,{x_2} + 
  0.0142072\,{{x_2}}^2 
\nonumber\\&&
 + 0.717803\,{x_3} - 
  0.367487\,{x_1}\,{x_3} - 0.48743\,{x_2}\,{x_3} 
\nonumber\\&&
+  0.29931\,{{x_3}}^2 
 - 0.770174\,{x_4} + 
  0.242792\,{x_1}\,{x_4} 
\nonumber\\&&
 - 0.0811259\,{x_2}\,{x_4} - 
  0.0954582\,{x_3}\,{x_4} + 0.113668\,{{x_4}}^2 
\nonumber\\&&
 + 0.460929\,{x_5} - 0.114934\,{x_1}\,{x_5} + 
  0.490873\,{x_2}\,{x_5} 
\nonumber\\&&
  - 0.691954\,{x_3}\,{x_5} + 
  0.588787\,{x_4}\,{x_5} - 0.810968\,{{x_5}}^2 
\nonumber\\&&
 - 3.27367\,{x_6} - 0.214978\,{x_1}\,{x_6} + 
  0.674502\,{x_2}\,{x_6} 
\nonumber\\&&
- 0.779896\,{x_3}\,{x_6} + 
  0.0891919\,{x_4}\,{x_6} + 1.37719\,{x_5}\,{x_6} 
\nonumber\\&&
+ 2.30296\,{{x_6}}^2 + 1.68227\,{x_7} - 
  0.289592\,{x_1}\,{x_7} 
\nonumber\\&&
- 0.19047\,{x_2}\,{x_7} - 
  1.18196\,{x_3}\,{x_7} - 0.0281177\,{x_4}\,{x_7} 
\nonumber\\&&
+ 
  0.517042\,{x_5}\,{x_7} + 0.67596\,{x_6}\,{x_7} - 
  0.454039\,{{x_7}}^2
\end{eqnarray}

\emph{Relation of this fit to those used in paper}:
The fit presented above is substantially less accurate than those actually used in our paper or in OKB: it is accurate only to within a factor
$1.8^{\pm 1}$ (i.e., when we evaluate this fit at all of our trial
points, we find the standard deviation between our results and the fit
to be $\left< (\log {\cal R}_{BH-BH,fit} -
\log{\cal R}_{BH-BH} )^2 \right>^{1/2} = 0.26$). 
The fits actually used in this paper are typically cubic (120 parameters) and quartic (330 parameters) order. Given the large number of these parameters we chose to just provide the quadratic-order fit as an example above.  However, the authors are happy to provide the  much longer expressions for the higher-order fits used upon request by any reader.

The rate functions are demonstrably \emph{not} separable: we cannot fit the rate functions well with a function of form $X_1(x_1) X_2 (x_2) \ldots X_7 (x_7)$.  Even this toy fit contains strong off-diagonal terms.

\section{C.~Characterizing the consistent region of population synthesis models}
\label{ap:describeregion}
\begin{figure}
\label{fig:allok}
\begin{center}
\includegraphics{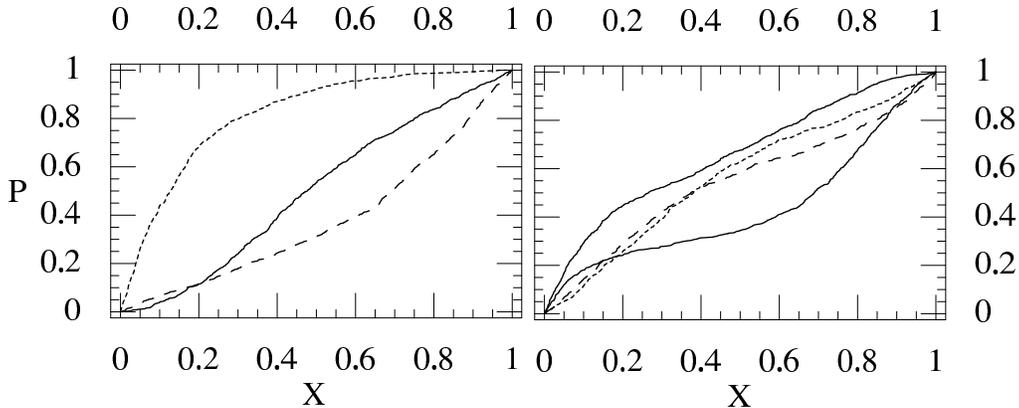}
\caption{Cumulative probability distributions
  $P_k(X)$ defined so $P_k(X)$ is the fraction of all models
  consistent with both constraints with
  $x_k<X$.  The left panel shows the distributions for the 3
  kick-related parameters $x_3,x_4,x_5$; in this panel, the bottom (dashed)
  curve denotes $P_3$, the middle (solid) curve denotes $P_4$, and the
  top dotted curve denotes $P_5$.  The right panel shows the
  distributions for $P_1$ (solid top line), $P_2$ (dashed), $P_6$
  (dotted), and $P_7$ (solid, bottom curve).
}
\end{center}
\end{figure}

Using our monte-carlo method to select models compatible with our
constraints, we have found a relatively small \emph{seven-dimensional
volume} consistent with observations that corresponds to about 2\% of all the runs we performed.   Unfortunately -- with some exceptions -- this volumetric
constraint does not translate to easily-understood and strong
constraints on the individual parameters $x_k$ (using the notation of
Appendix~\ref{ap:samplefits}).  On the one hand, because the dimension
is high, weak constraints on each parameter can correspond to very
strong volumetric constraints.  On the other hand, as demonstrated in Figure~\ref{fig:allok}7, because the consistent region is extended through our high-dimensional model space in a
inhomogeneous anisotropic fashion, \emph{wide ranges of 
values} of each parameter are still allowed, even after applying the
constraints.

To provide the reader with a global view of the parameter values associated with the models that turn out to be consistent with our constraints, in Figure~\ref{fig:allok}7 we show the cumulative distributions of the consistent model parameter values for each of the seven parameters. 
 It is evident that the full ranges of values ([0,1]) are covered by the model parameters $x_{k}$ for the set of models consistent with the constraints. However, certain qualitative conclusions can be drawn:
about 80\% of the consistent models have kick relative weights (parameter $x_{5}$) below 0.3; about 50\% of the consistent models have mass-ratio power-law indices smaller (in absolute value) than 0.6 ($x_{1}<0.2$; fractions of mass lost from the binary during non-conservative mass transfer phases in the range 20\%-60\% are not favored.  

Given the above, any simple attempt to describe the consistent region will necessarily be a crude approximation.  Nonetheless, for completeness we  attempt to
characterize the extended consistent region through its mean values.  
The mean model consistent with our constraints is given by: 
\begin{equation}
\bar{x} = (x_1,x_2,\ldots, x_7)
= (0.33,\; 0.46, \; 0.61,\; 0.53,\; 0.18,\; 0.43,\; 0.57)
\end{equation}
These mean values correspond to a model with: fairly flat mass
ratio distribution (power-law of $\simeq -1$); moderate stellar winds
(strengths reduced by factors of $\simeq 2$); moderate kicks drawn
from Mawellians with $\sigma_{1}\simeq 120$\,km\,s$^{-1}$,
$\sigma_{2}\simeq 625$\,km\,s$^{-1}$, and with relative weights of
$\simeq 20$\% (favoring $\sigma_2$ over $\sigma_1$); moderate values
for an effective common-envelope efficiency ($\alpha\lambda \simeq
0.4$ including the central concentration parameter $\lambda$); and
moderately non-conservative mass transfer phases ($\simeq 40$\% of the mass is lost from the binary). 
 It is very important though to keep in mind the broad ranges of these parameters shown in Figure~\ref{fig:allok}7. 


\end{document}